\documentclass{ws-ijmpa}
\begin{document}

\markboth{G.~L.~Klimchitskaya  {\it et al.}}
{Observation of the Thermal Casimir Force Is Open
to Question}

%
\catchline{}{}{}{}{}
%

\title{OBSERVATION OF THE THERMAL CASIMIR FORCE IS OPEN
TO QUESTION}

\author{
G.~L.~KLIMCHITSKAYA,${}^{1,2}$ M.~BORDAG,${}^{3}$
E.~FISCHBACH,${}^{4}$
D.~E.~KRAUSE${}^{5,4}$ and V.~M.~MOSTEPANENKO${}^{1,6}$
}

\address{${}^1$Department of Physics, Federal University of Para\'{\i}ba,
C.P. 5008, CEP 58059--900,{\protect \\} Jo\~{a}o Pessoa, Pb-Brazil{\protect \\}
${}^2$North-West Technical University,
 Millionnaya Street 5, St.Petersburg,
191065, Russia{\protect \\}
${}^3$Institute for Theoretical
Physics, Leipzig University, Postfach 100920,
D-04009, Leipzig, Germany
{\protect \\}
${}^4$Department of Physics, Purdue University, West Lafayette, Indiana
47907, USA{\protect \\}
${}^5$Physics Department, Wabash College, Crawfordsville, Indiana 47933,
USA
{\protect \\}
${}^6$Noncommercial Partnership
``Scientific Instruments'',
Tverskaya Street 11, Moscow,
103905, Russia
}

\maketitle

\begin{history}
\received{2 June 2011}
\end{history}

\begin{abstract}
We discuss theoretical predictions for the thermal Casimir force
and compare them with available experimental data.
Special attention is paid to the recent claim of the observation
of that effect, as predicted by the Drude model approach.
We show that this claim is in contradiction with a number of
experiments reported so far. We suggest that
the experimental errors, as reported in support of the observation
of the thermal Casimir force, are significantly underestimated.
Furthermore, the experimental data at separations above $3\,\mu$m
are shown to be in agreement not with the Drude model approach,
as is claimed, but with the plasma model.
The seeming agreement of the data with the Drude model at
separations below $3\,\mu$m is explained by the use of an
inadequate formulation of the proximity force approximation.

\keywords{Casimir force; thermal corrections; precise measurements.}
\end{abstract}

\ccode{PACS numbers: 12.20.-m, 42.50.Ct, 78.20.Ci}

\section{Introduction}

During the last ten years much attention has been given to the Casimir
force at nonzero temperature. This physical phenomenon is described
by the Lifshitz theory\cite{1} which presents the Casimir free
energy and force between two parallel plates as a functional of
the dielectric permittivity of plate materials calculated along
the imaginary frequency axis, $\varepsilon(i\xi)$.
The optical data for the complex index of refraction\cite{2}
extrapolated to low frequencies allow the calculation of
$\varepsilon(i\xi)$ using the Kramers-Kronig relation.
Surprisingly, for metal test bodies the use of most natural
extrapolation by means of the Drude model was shown to be
in violation of the Nernst heat theorem\cite{3,4} and
in contradiction with experimental
data.\cite{5}\cdash\cite{8}
On the other hand, the extrapolation of the optical data
below the edge of the absorption bands by means of the
plasma model, which disregards dissipation of conduction
electrons, turned out to be in agreement
with the Nernst theorem, and
 consistent with the experimental results.
This created a serious problem because in accordance with
classical Maxwell equations the dielectric permittivity
in the quasistatic regime is inversely proportional to
the frequency in accordance with the Drude model,
whereas the plasma model is an approximation
applicable only at sufficiently high (infrared)
frequencies. Currently the use of the Drude and plasma
models in the Lifshitz formula is customarily called
the Drude\cite{9}\cdash\cite{11} and
plasma\cite{8,12}\cdash\cite{14} model approaches,
respectively.

In this paper we discuss present experimental status of
the Drude and plasma model approaches to the thermal
Casimir force. In Sec.~2 main characteristic features
of measurements of the Casimir pressure between metal
test bodies by means of a micromechanical oscillator
are considered. These measurements exclude the Drude
model approach but are consistent with the plasma
model. Section~3 briefly reviews experiments with
semiconductor\cite{15,16} and dielectric\cite{17,18}
test bodies, where the Drude model approach was
excluded as a description of the dc conductivity
of dielectrics.
In Sec.~4 experiments\cite{19}\cdash\cite{21}
using a torsion balance and spherical lenses with
centimeter-size curvature radii performed before
2010 are discussed. The first of them\cite{19}
was interpreted\cite{22} as being in disagreement
with the Drude model, but later this conclusion
was cast in doubt.\cite{23}
Our main attention here is devoted to the recent
experiment\cite{24} claiming the observation of
the thermal Casimir force, as predicted by the
Drude model approach (see Sec.~5).
We demonstrate that at separations below
$3\,\mu$m the interpretation of this experiment
is in fact uncertain because  surface imperfections
were ignored, and these
are invariably present on surfaces of
lenses of centimeter-size curvature
radius.\cite{24a} At separations above
$3\,\mu$m the experimental data are shown to
agree with the plasma model approach, as opposed
to what is claimed in Ref.~\refcite{24}.
Section~6 contains our conclusions and discussion.

\section{Experiments Between Metal Test Bodies Using
a Micromachined Oscillator}

These three experiments conducted with increased
precision\cite{5}\cdash\cite{8} are independent
measurements of the gradient of the Casimir force acting
between a sphere of 300 or $150\,\mu$m radius and a
plate of a micromechanical torsional oscillator,
both covered with Au. Using the proximity force
approximation (PFA), which leads to negligibly small
errors for perfectly spherical surfaces of sufficiently large
curvature
radii, the gradient of the Casimir force was reexpressed
as the Casimir pressure between two
parallel plates. Different voltages were applied between
the sphere and the plate, and the electric force was measured
and found in agreement with exact theoretical results in a
sphere-plate geometry. This was used to perform
electrostatic calibrations. Specifically, the residual
potential difference between the sphere and the plate in
the absence of applied voltages was determined and found to
be independent of separation. (The details of these
experiments are described in Refs.~\refcite{25}
and \refcite{26}.)

It should be stressed that the third, and most precise,
experiment using a micromachined oscillator\cite{7,8}
is of metrological quality, because the random error
of the measured Casimir pressures was made much
smaller than the systematic (instrumental) error.
The comparison of the measurement results with different
theoretical approaches was performed taking into account
all possible undesirable systematic effects. Thus, the role of
patch potentials was investigated\cite{6} and found to be
negligibly small (here only small patches are
possible with a maximum size less than the thickness
of the Au layer, i.e., less than 300\,nm).

\begin{figure*}[t]
\vspace*{-7.8cm}
\centerline{\hspace*{4.cm}\psfig{file=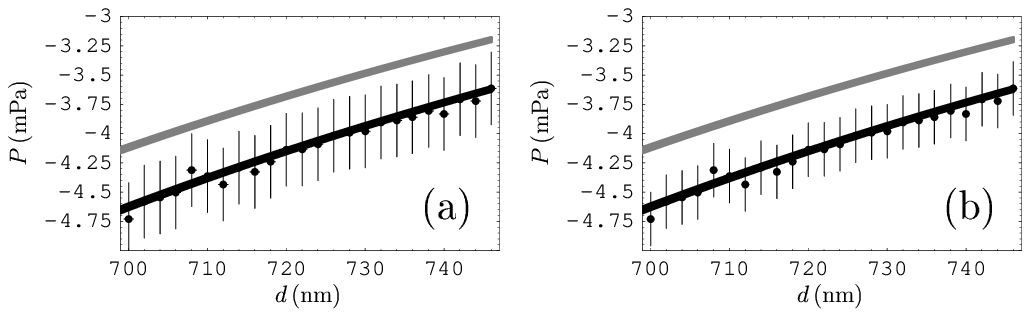,width=24cm}}
\vspace*{-22.4cm}
\caption{The experimental data for the Casimir
pressure between two parallel plates measured${}^{7,8}$
by means of micromechanical torsional oscillator as a function
of separation are shown as crosses.
The arms of the crosses indicate the total experimental
errors  determined at
(a) 95\%  and (b) 70\% confidence level.
The grey and black lines show
the theoretical Casimir pressures computed
 using the Drude and plasma model approaches,
 respectively.
 The thickness of the lines indicates the total
 theoretical errors.}
\end{figure*}

The experimental results exclude the predictions of the Drude
model approach over the entire measurement range from
162 to 746\,nm at a 95\% confidence level. The same
results were found to be consistent with the plasma model
approach. For the purpose of comparison with the large
separation experiment\cite{24} in Sec.~5, in Fig.~1 we
present the measurement data indicated as crosses in comparison
with theoretical predictions in the region from 700 to 746\,nm.
In Fig.~1(a) and 1(b) the arms of the crosses indicate the
total experimental errors determined at a
95\% and 70\% confidence levels, respectively.
As can be seen in Fig.~1, the prediction of the plasma model
approach (shown by the black line) is in excellent agreement
with the data, whereas the prediction of the Drude model
(the grey line) is experimentally excluded.
It should be stressed that
the experimental and theoretical results shown in Fig.~1
are independent, and the comparison between experiment and
theory has been performed with no fitting parameters.

\section{Experiments with Semiconductor and Dielectric
Test Bodies}

The Drude-type dielectric permittivity is also used to describe
dc conductivity in dielectrics and semiconductors of dielectric
type. It is traditional\cite{1} to disregard
dc conductivity when dealing with the van der Waals and Casimir
forces between dielectric test bodies. This was justified by the
presumed smallness of this effect. It was shown,\cite{27}
however, that the inclusion of dc conductivity into the Lifshitz
theory leads to a violation of the Nernst heat theorem and
significantly increases the magnitude of the Casimir force.
This effect was tested in the experiment\cite{15,16}
measuring the Casimir force difference between an Au sphere of
$100\,\mu$m radius and a Si plate illuminated with laser pulses
performed using an atomic force microscope.
In the absence of laser light, the Si plate was in a dielectric
state with a density of free charge carriers
$5\times 10^{14}\,\mbox{cm}^{-3}$. In the presence of light,
the density of free charge carriers was increased
by almost 5 orders
of magnitude (semiconductor of metallic type).
The experimental results for the Casimir force difference
$F_C^{\,\rm diff}$ (in the presence minus in the absence
of light)
exclude the dc conductivity described by the Drude model at a
95\% confidence level, but are consistent with the Lifshitz
theory with dc conductivity in the absence of light disregarded.
In Fig.~2(a) the experimental data for the difference Casimir
force  are indicated as crosses.
The theoretical results are shown by the black and grey lines
with dc conductivity in the dark phase disregarded and included,
respectively. In this experiment, the experimental and
theoretical results were also obtained independently, and
their comparison has been made with no fitting parameters
(see Refs.~\refcite{25} and \refcite{26} for details).
\begin{figure*}[t]
\vspace*{-7.7cm}
\centerline{\hspace*{4.cm}\psfig{file=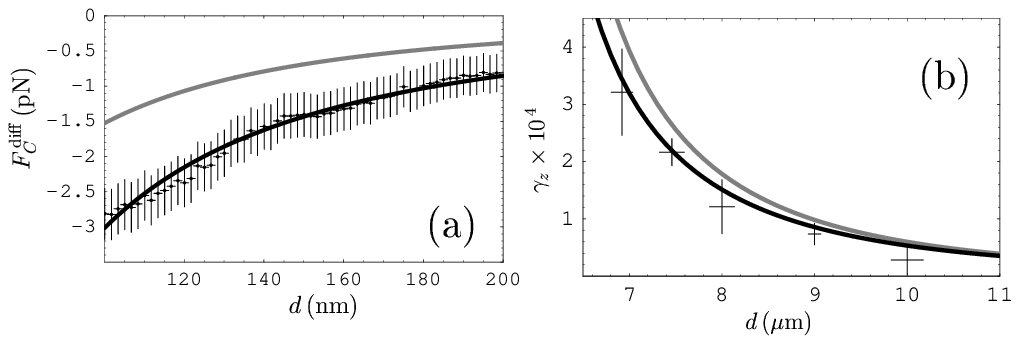,width=24cm}}
\vspace*{-22.4cm}
\caption{(a) The experimental data for the difference Casimir
force between an Au sphere and Si plate measured${}^{15,16}$
by means of AFM as a function of separation are shown as
crosses.
The arms of the crosses indicate the total experimental errors
determined at a 95\% confidence level.
The black and grey lines show
the theoretical difference forces computed
with neglected and included dc conductivity of Si plate
in the absence of light, respectively.
(b) The experimental data for the fractional change of the trap
frequency due to the Casimir-Polder force between ${}^{87}$Rb
atoms and fused silica plate are indicated by crosses
as a function of separation.
The total experimental errors are
determined at a 70\% confidence level.
The black and grey lines show the theoretical fractional
shift computed with neglected
and included dc conductivity of fused silica.
}
\end{figure*}

Another important experiment is the measurement of the thermal
Casimir-Polder force between ground state ${}^{87}$Rb atoms,
belonging to a Bose-Einstein condensate, and a fused silica
plate at separations from about 7 to $10\,\mu$m.
This experiment was repeated three times, in equilibrium,
when the plate temperature was the same as that of environment,
and out of equilibrium,
when the plate temperature was higher than in the environment.
In all cases  the measurement
data were in agreement with theory disregarding dc
conductivity of fused silica,\cite{17} but were in
contradiction with theory taking this conductivity into
account.\cite{18} As an example, Fig.~2(b) shows the
measured fractional shift of the trap frequency
$\gamma_z$ (crosses) due to the Casimir-Polder force as
a function of separation. The arms of the crosses are
plotted at a 70\% confidence level (an environment was at
310\,K and the plate at 605\,K). The black and grey
lines show theoretical results computed disregarding
and including dc conductivity of fused silica, respectively.
We emphasize that, as in the previous two cases, this
experiment is an independent measurement and no fitting
parameters have been  used when comparing  the data
with theory.

\section{Torsion Balance Experiments}

Experiments measuring the Casimir force with a torsion
pendulum\cite{19}\cdash\cite{21,24} use the configuration of a
spherical lens of greater than 10\,cm radius of curvature $R$ in
close proximity to a plate. The first such experiment\cite{19} was
performed in 1997 with a lens of $R=12.5\,$cm and a plate both
coated with Au, and was criticized\cite{28,29} for overestimation
of the level of agreement between the measurement data and
theory. The results of this experiment at about $1\,\mu$m
separation were used\cite{22} to exclude the Drude model
approach to the Casimir force. At $d=1\,\mu$m the latter
predicts a --18.9\% thermal correction to the force which was
not observed. Recently the possibility of a systematic
correction due to time-dependent fluctuations in the distance
between the lens and the plate was discussed.\cite{23}
It was speculated\cite{23} that such a correction, if it is
relevant to the experiment of Ref.~\refcite{19}, might bring
the data into agreement with the Drude model approach.

Another torsion balance experiment\cite{20} used a lens of
$R=20.7\,$cm curvature radius and a plate also coated with Au.
The measured data over the separation region from 0.48 to
$6.5\,\mu$m demonstrated a high level of agreement
with the standard theory of the Casimir force and did not
support the existence of large thermal corrections predicted
by the Drude model. The comparison between the
data and the standard theory of the Casimir force in
Ref.~\refcite{20} is characterized by the $\chi^2=513$ with
the number of degrees of freedom equal to 558.
{}From this it follows\cite{33} that the probability of
obtaining a larger value of the reduced $\chi^2$ in the
next individual measurement is as large as 91\%,
that is a high level of agreement.

One more torsion balance experiment\cite{21} used a Ge lens
and a Ge plate. The experimental precision of this experiment
was not sufficient to discriminate between different
theoretical approaches to the Casimir force.

The characteristic feature of the torsion pendulum experiments
listed above is that all of them use fitting parameters,
 such as the force offset, the offset of the voltage
 describing a noncompensated electric force, etc.
 The values of these parameters are found from the best fit
 between the experimental data and different theories.
 This means that torsion balance experiments are not
 independent measurements, and the comparison of their
 results with theory is not as definitive as for the
 independent measurements considered in Secs.~2 and 3.

 An advantage of torsion pendulum experiments is the use
 of lenses with centimeter-size radii of curvature, which
 significantly increases the magnitude of the Casimir force
 and allows measurements at separations of a few micrometers.
 The use of large lenses, however, leads to a problem:
 the experimental results in
 Refs.~\refcite{19}--\refcite{21} were compared with theory
 using the simplest version of the PFA for the
 Casimir force acting between a lens and a plate\cite{25}
 \begin{equation}
 F(d,T)=2\pi R{\cal F}(d,T),
 \label{eq1}
 \end{equation}
 \noindent
 where ${\cal F}(d,T)$ is the free energy (per unit area)
 at temperature
 $T$  in the configuration of two parallel
 plates. Recently it was shown\cite{30} that for lenses
 of centimeter-size curvature radius at separations below
a few micrometers, Eq.~(\ref{eq1}) is not applicable due
to deviations from perfect sphericity (such as bubbles and
pits) which are invariably present on lens
surfaces.\cite{24a} For example, if there is a bubble
with the radius of curvature $R_1$ and thickness $D$ near
the point of closest approach to the plate, the general
formulation of the PFA\cite{25} leads not to Eq.~(\ref{eq1}) but
to the result:\cite{30}
 \begin{equation}
 F(d,T)=2\pi (R-R_1){\cal F}(d+D,T)+2\pi R_1{\cal F}(d,T).
 \label{eq2}
 \end{equation}
 \noindent
Calculations show\cite{30} that the Casimir force between
a perfectly spherical lens and a plate described by the
Drude model at $d<3\,\mu$m can be made approximately
equal to the force between a lens with some surface
imperfections and a plate described by the plasma model,
and vice versa. This makes measurements of
the Casimir force by means of torsion pendulum
experiments uncertain at separations below $3\,\mu$m.

\section{Purported Observation of the Thermal
Casimir Force}

Recently, one more experiment measuring the thermal
Casimir force has been performed\cite{24} using the
torsion pendulum technique. The attractive force between
an Au-coated spherical lens of $R=(15.6\pm 0.31)\,$cm
radius of curvature and an Au-coated plate was measured
over a wide range of separations from 0.7 to $7.3\,\mu$m.
As in the case for the
earlier torsion pendulum experiments mentioned
in Sec.~4, the experiment of Ref.~\refcite{24} is not
an independent measurement. It uses two phenomenological
parameters determined from the best fit between the
experimental data and different theoretical approaches
(see below for details).
Furthermore, similar to earlier experiments exploiting
large spherical lenses, the experiment of Ref.~\refcite{24}
ignores surface imperfections that are invariably
present on surfaces of real lenses, as discussed in
Sec.~4, and uses Eq.~(\ref{eq1}) in computations
notwithstanding the fact that it
is applicable only for perfectly spherical
surfaces.

It should be emphasized that the experiment of
Ref.~\refcite{24} measures not the thermal Casimir force
in itself, but up to an order of magnitude larger total
attractive force between a lens and a plate.
The total force is assumed to be the sum of the Casimir
force and the electrostatic force from
 the large  patches.
As the authors themselves recognize,\cite{24}
``an independent measurement of this electrostatic
force with the required accuracy is currently not
feasible''. That is why it is hypothesized that there
are large patches on Au-coated surfaces due to
absorbed impurities or oxides whose size $\lambda$
satisfies the condition
\begin{equation}
d\ll\lambda \ll r_{\rm eff}=\sqrt{Rd}.
\label{eq3}
\end{equation}
\noindent
Note that small patches due to spatial changes in
surface crystalline structure, also mentioned in
Ref.~\refcite{24}, satisfy a condition
$\lambda\ll d$ because the thickness of the Au
films used is only 70\,nm. The electric force due to
such small patches is exponentially small.\cite{31}
Keeping in mind the parameters of the configuration
used in Ref.~\refcite{24}, for the size of large patches
satisfying Eq.~(\ref{eq3}) one obtains
$\lambda\approx 50\,\mu$m.
The electric force due to large patches was modelled
by the term\cite{24}
\begin{equation}
F_{\rm patch}(d)=-\pi\epsilon_0 R\frac{V_{\rm rms}^2}{d},
\label{eq4}
\end{equation}
\noindent
where $\epsilon_0$ is the permittivity of free space, and
the parameter $V_{\rm rms}$ describes the magnitude of the
voltage fluctuations across the test bodies.
The value of this parameter was not measured and it was
used as one of the fitting parameters (we assume that an
attractive force is negative).

In the presence of a voltage $V$ applied between the lens
and the plate the measured force was represented in the
form\cite{24}
\begin{equation}
F(d,V)=F_C(d)-\pi\epsilon_0 R\left[
\frac{(V-V_{m})^2}{d}+
\frac{V_{\rm rms}^2}{d}\right],
\label{eq5}
\end{equation}
\noindent
where $V_m\approx 20\,$mV is the residual potential
difference between the lens and the plate and
$F_C(d)$ is the Casimir force at laboratory temperature
$T=300\,$K. It was found\cite{24} that $V_m$ is nearly
independent of separation where it was determined
(the variation was 0.2\,mV between 0.7 and $7\,\mu$m).

The measurement data for the total force at different
separations were taken with an applied potential equal to
$V_m$ in order to cancel
the force from the residual potential difference.
The data were corrected for the presence of fluctuations
in separation and for a long-term drift of a
vibration-isolation slab. The mean experimental data
for the total measured force (obtained from 383 sweeps)
multiplied by separations are shown on a logarithmic
scale in Fig.~2 of Ref.~\refcite{24} together with their
errors  as a function of separation.
We reproduce these data in our Fig.~3 plotted on a linear scale.
\begin{figure*}[t]
\vspace*{-11.1cm}
\centerline{\hspace*{4.cm}\psfig{file=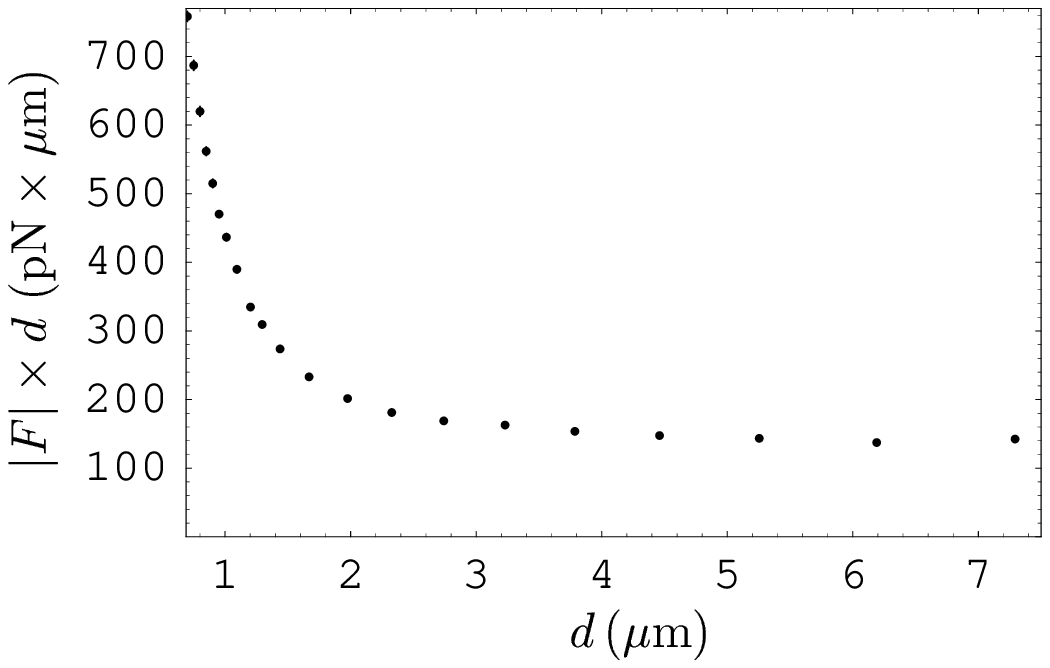,width=15cm}}
\vspace*{-5.cm}
\caption{The experimental data for the magnitude of the mean
measured force${}^{24}$ (electric plus Casimir) multiplied by
separation are shown as crosses. The arms of the crosses
indicate the experimental errors.${}^{24}$
}
\end{figure*}
As can be seen in Fig.~3, the errors are unexpectedly small.
For example, at the largest separation, $d=7.29\,\mu$m,
the measured total force is equal to
$F=(19.54\pm 0.28)\,$pN, leading to the relative error 1.4\%.
For the remaining 20 data points the relative errors vary from
0.86\% to 2.2\% (note that according to Ref.~\refcite{24} the
total force ``is measured at 30 logarithmically spaced plate
separations," but for some unexplained reason the data at
only 21 separations are shown\cite{24} in Figs.~2 and 3).
According to the caption to Fig.~2 in Ref.~\refcite{24},
``the vertical error bars include contributions from the
statistical scatter of the points as well as from uncertainties
in the applied corrections.'' Such an important contribution
of the total experimental error as a systematic (instrumental)
error, which is understood as an error of a calibrated device used
in force measurements (i.e., the smallest fractional division of
the scale of the device), is not mentioned.
The omission of a systematic error may explain the claimed
smallness of all errors in Figs.~2 and 3 of Ref.~\refcite{24}.
The point is that the absolute instrumental errors are
typically constant, and this leads to a quick increase of the
total relative error with increasing separation distance.
If this is true here, it calls into question
the subsequent analysis and conclusions
in Ref.~\refcite{24}. Below, however, we attempt to follow the
line of reasoning in Ref.~\refcite{24}  as if the errors
indicated there in Fig.~2 were the total experimental errors.
One should also mention that the confidence level at which the
errors are found is not indicated in Ref.~\refcite{24}.
We assume that it is on the level of one sigma.

The corrected mean data were fitted to the theoretical
expression of the form\cite{24}
\begin{equation}
F(d)=F_C(d)-\pi\epsilon_0 R\frac{V_{\rm rms}^2}{d}-a,
\label{eq6}
\end{equation}
\noindent
where in comparison with (\ref{eq5}) one more fitting parameter
$a$ was introduced  as a constant force offset due to
voltage offsets in the measurement electronics.\cite{24}
Here, $F_C(d)$ is the theoretical thermal Casimir force which
can be computed in the framework of the Lifshitz theory\cite{1}
using either the Drude or the plasma model approach.
{}From the fitting of mean experimental data for the total force
to the theoretical force in Eq.~(\ref{eq6}) with two fitting
parameters $V_{\rm rms}$ and $a$, it was concluded
in Ref.~\refcite{24} that the
data are in excellent agreement with the thermal Casimir force
calculated using the Drude model approach. The plasma model
was excluded in the measured separation range. Below we
demonstrate that these conclusions are in fact not supported.

Computations of the thermal Casimir force within the Drude model
approach were performed in Ref.~\refcite{24}
using the tabulated\cite{32} optical
data for the complex index of refraction of Au extrapolated to
low frequencies by means of the Drude model. The claimed excellent
agreement of these computations with the data manifested itself as
the value of reduced $\chi^2=1.04$ with the fitting parameters
$a=-3.0\,$pN and $V_{\rm rms}=5.4\,$mV.
It is easily seen, however, that
in the experiment under consideration
(the number of degrees of freedom is equal to $21-2=19$)
this value of the reduced $\chi^2$
implies that
the probability of obtaining a larger value
of the reduced $\chi^2$ in the next individual measurement is
equal\cite{33} to 41\%.
For the results of an individual measurement fitted to some
model, such a $\chi^2$-probability could be considered as
being in favor of this model. If, however, the mean
measured force over a large number of repetitions is used
for the fit, as in Ref.~\refcite{24}, the $\chi^2$-probability
should be larger than 50\% in order the measured data could be
considered as supporting the theoretical model.

The next important fact  is that at separations
$d>3\,\mu$m the experimental data of Ref.~\refcite{24} are
not in agreement with the Drude model approach, as is claimed,
but with the plasma model. At such large separations the
theoretical predictions of the Drude and plasma model
approaches to a large extent do not depend on the values of
plasma frequency and relaxation parameter.
Furthermore, at $d>5\,\mu$m the Casimir forces calculated
using the Drude and plasma model approaches are
approximately given by\cite{25,26}
\begin{equation}
F_C^D(d)=-\frac{\zeta(3)Rk_BT}{8d^2},\qquad
F_C^p(d)=-\frac{\zeta(3)Rk_BT}{4d^2},
\label{eq7}
\end{equation}
\noindent
where $\zeta(z)$ is the Riemann zeta function, i.e., the
predictions of both approaches differ by a factor of two.

To compare the experimental data for the total force with
the Drude and plasma model approaches at $d>3\,\mu$m (six
experimental points in Fig.~2 of Ref.~\refcite{24} and
in our Fig.~3),
we have repeated the fitting procedure to Eq.~(\ref{eq6})
with all the same corrections as were introduced by the authors.
Keeping in mind that Eq.~(\ref{eq6}) contains two fitting
parameters, the number of degrees of freedom is
equal to four.
When the Drude model approach at $T=300\,$K is used in the
fit, the best agreement with Eq.~(\ref{eq6}) is achieved
at $a=-0.29\,$pN and $V_{\rm rms}=5.45\,$mV.
The respective reduced $\chi^2=1.65$ leads\cite{33}
to the probability of obtaining a larger reduced $\chi^2$
in the next measurement equal to 16\%.
This signifies a poor agreement of the data with
the predictions of the Drude model.

To check the predictions of the plasma model approach at
$d>3\,\mu$m, we used the generalized plasma-like
model\cite{8,14,25,26} to calculate the thermal Casimir
force. In this case at $T=300\,$K
the best agreement between Eq.~(\ref{eq6})
and the data is achieved with
$a=3.6\,$pN, $V_{\rm rms}=4.5\,$mV and the
reduced $\chi^2=0.67$. Thus, in the next measurement
 a larger reduced $\chi^2$ will be obtained\cite{33}
with the probability 61\%. This confirms that the
plasma model is in a good agreement with the
large separation data of Ref.~\refcite{24}.
\begin{figure*}[t]
\vspace*{-11.1cm}
\centerline{\hspace*{4.cm}\psfig{file=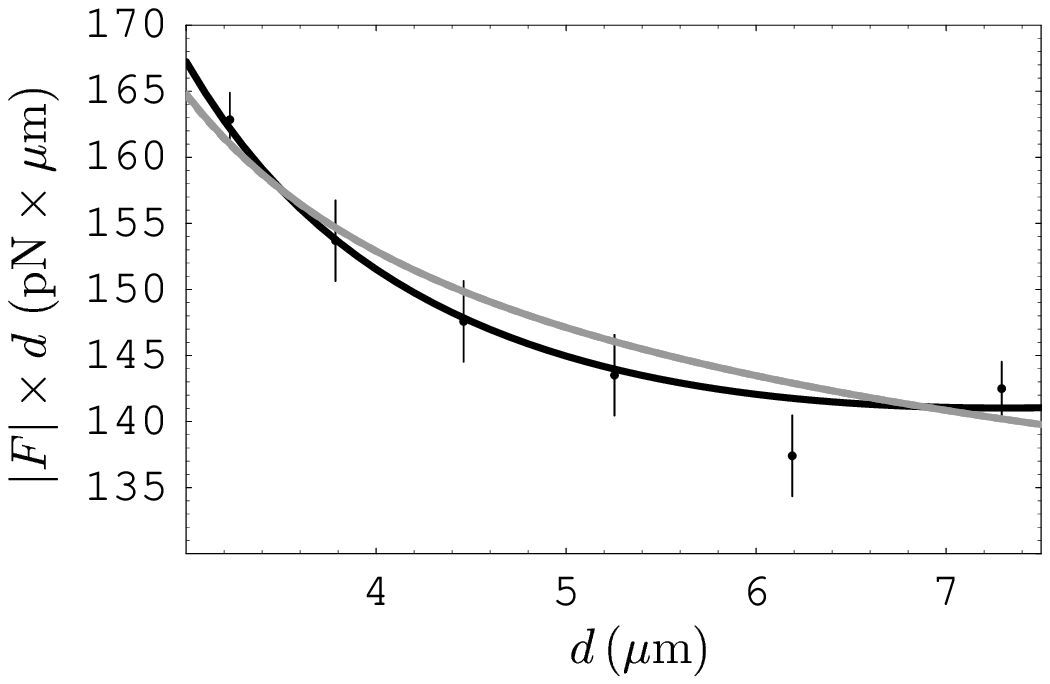,width=15cm}}
\vspace*{-5.cm}
\caption{The experimental data for the magnitude of the mean
measured force${}^{24}$  multiplied by
separation are show as crosses. The arms of the crosses
indicate the experimental errors.${}^{24}$
The grey and black lines demonstrate the best fit to the
experimental data of the total theoretical force (electric
plus Casimir) computed using the Drude and plasma model
approaches, respectively, with two fitting parameters.
}
\end{figure*}
In Fig.~4 the magnitudes of the total theoretical forces
(electric plus Casimir) multiplied by separations are
shown by the grey and black lines.
They are obtained from the best fit to the experimental
data of Ref.~\refcite{24}, indicated as crosses, using
the Drude and plasma model approaches, respectively.
It is seen that the force data at $d>3\,\mu$m are in
a good agreement with the plasma model approach.

A seeming agreement of the fit performed in
 Ref.~\refcite{24} with the Drude model approach at
 separations $d<3\,\mu$m may be explained by an
 unjustified use of the PFA in the simplest form of
 Eq.~(\ref{eq1}). As explained in Sec.~4, this
 formulation of the PFA is not applicable to the
 configuration of a large lens and a plate spaced
 at separations below $3\,\mu$m. The surfaces of lenses with
 large radius of curvature ($R=15.6\,$cm in
  Ref.~\refcite{24}) are characterized by deviations from
  perfect sphericity which necessitate
  the use of more sophisticated
  formulations of the PFA, such as in Eq.~(\ref{eq2}).
It was shown\cite{30} that in such situations, the force with the
plasma model incorporating surface imperfections can appear at
separations below $3\,\mu$m as if due to the Drude model.

\section{Conclusions and Discussion}

In the foregoing, we have discussed the comparison between
experiment and theory in connection with claimed observation
of the thermal Casimir force
 in Ref.~\refcite{24}, as predicted by the Drude model
approach.
Our first conclusion is that the experimental results of
Ref.~\refcite{24} are in contradiction with a number of
experiments which exclude this theoretical description of
the thermal Casimir force (such as the three experiments
using a micromachined oscillator,\cite{5}\cdash\cite{8}
two experiments using
 a torsion pendulum,\cite{19,20} an experiment on optical
modulation of the Casimir force\cite{15,16} performed using
an atomic force microscope, and an experiment on measuring
the Casimir-Polder force\cite{17,18}).

Our second conclusion is that the experiment\cite{24} is not
an independent measurement of the thermal Casimir force,
but a fit using two fitting parameters and a phenomenological
expression for the electric force, presumably arising from
large surface patches. According to the authors, this
electric force cannot be measured with sufficient precision,
although it is up to an order of magnitude greater than the
thermal Casimir force. Keeping in mind that the
experiments\cite{5}\cdash\cite{8,15}\cdash\cite{17} are independent
measurements of the Casimir and Casimir-Polder force or the
Casimir pressure, one may cast doubt on the results of the
experiment of Ref.~\refcite{24}.
Note also that experiments\cite{19,20} use a fitting procedure
similar to Ref.~\refcite{24}, but arrive at the opposite
conclusion that the Drude model is not supported.

According to our third conclusion, the experimental errors
in the mean total force in Ref.~\refcite{24} are significantly
underestimated. This is seen from the fact that the
systematic (instrumental) errors, which are typically
separation-independent, were not addressed and
taken into account in the balance of errors.
This resulted in surprisingly small relative experimental
errors in the mean measured forces (equal to, e.g.,
1.4\% at the separation $7.29\,\mu$m).

The fourth conclusion is that at separations above $3\,\mu$m
the measurement data\cite{24} are in agreement not with the
Drude model approach to the thermal Casimir force, as is
claimed,\cite{24} but with the plasma model.
This is readily demonstrated by the application of the
fitting procedure\cite{24} with all the same corrections,
as made by the authors of Ref.~\refcite{24}, to the last
six experimental points measured at separations
above $3\,\mu$m.

The last, fifth, conclusion is that a seeming agreement of
the experimental data for the mean total force\cite{24}
with the Drude model at separations below $3\,\mu$m can be
explained by the use\cite{24} of an inadequate formulation of
the PFA. The formulation that was employed is applicable only to
perfectly shaped spherical surfaces, whereas lenses of
centimeter-size radius of curvature (such as exploited
in Ref.~\refcite{24}) invariably contain surface
imperfections that must be taken into account in
computations.

To conclude, the results of Ref.~\refcite{24} cannot be
considered as a reliable confirmation of the predictions
of the Lifshitz theory combined with the Drude model.
Therefore, the problem of thermal Casimir force remains
to be solved.

\section*{Acknowledgments}
G.L.K.\ and V.M.M.\ were supported by CNPq (Brazil) and by the DFG
grant BO\,\,1112/20--1.
E.F. was supported in part by DOE under Grant No.~DE-76ER071428.


\end{document}